\documentclass[a4paper]{article}
\usepackage[psamsfonts]{amssymb}
\usepackage[dvips,dvipdf]{graphicx}
\usepackage{amssymb}

\begin{document}

\title{\sc Transmission of a quantum state in a periodically poled nonlinear crystal}
\author{R. Singh \\
\\
House No. 32, Industrial Area, Panipat 132103, India. \\
Email: ranjit.singh@mail.ru}
\maketitle
\pagestyle{empty}\thispagestyle{empty}

\begin{abstract}
Theoretical calculations on transmission of quantum states such as Schr\"{o}dinger cat (SC) states are considered in a periodically poled nonlinear crystal (PPNC). Combinations of various initial states (SC, coherent (C),and vacuum (V)) of light waves at frequencies $\omega_{e}$ and $3\omega_{e}$ at the input of a PPNC are studied. It is shown that the transmission and interference between superposition of macroscopically distinguishable states can be achieved by using a PPNC. Visualization and analyzes of transmission and interference processes are demonstrated with the help of reduced Wigner quasi-probability distribution functions of considered light waves.
\end{abstract}

\section{Introduction}
In the recent years, PPNCs are frequently used in generation of nonclassical states of light and realization of quantum information protocols [1-4]. The main advantages of using PPNCs against ordinary nonlinear optical crystals are: the quasi-phase-matching condition between interacting waves; the highest nonlinear susceptibility coefficient can be used; multi-mode interaction of optical waves.\\

Over the last years researchers have been paying much attention to quantum state engineering. This is due to the realization of various quantum information protocols where quantum nonlinear optics plays a fundamental role in generation of various nonclassical states. These nonclassical states play an important role in implementation of quantum state teleportation, cryptography and imaging [1-5]. One of special case of nonclassical states in quantum nonlinear optics is SC state $(|\alpha>\pm |-\alpha>)/\sqrt{2\pm 2e^{-2|\alpha|^2}}$ [6,7]. SC state possesses quantum interference effects [7,9] between superposition of macroscopically distinguishable states $|\alpha>$ and $|-\alpha>$. This effect can be visualized by the application of Wigner quasi-probability distribution function [5-7,9]. A possible scheme of generation of SC state in an ordinary nonlinear optical crystal with second order nonlinear susceptibility was numerically studied in paper [8]. In this paper authors had shown the possibility of generation of SC state in the coherent fundamental pump light wave of frequency $\omega_{o}$ while second harmonic wave of frequency $2\omega_{e}$ at the input of an ordinary nonlinear optical crystal was in the vacuum state. Generation of SC appears by considering full quantum mechanical treatment of interaction Hamiltonian $\hbar(\xi \hat{a}_{1o}^{2}\hat{a}_{2e}^{+}+HC)$, i.e. without assumptions of any classical and non-depleted assumptions at the input of an ordinary nonlinear optical crystal. Unfortunately, the SC state destroys over long interaction length in an ordinary nonlinear crystal [8].\\

Expect generation of SC state there are also various other challenging tasks in quantum state engineering such as how to manipulate and transmit SC state while manipulating and preserving quantum interference effects. Here we will study the transmission and interference processes of SC state from light wave of frequency $\omega_{e}$ to $3\omega_{e}$ and vice-versa in a PPNC.\\

The paper is structured as follows. Section 2 describes the optical nonlinear processes and their Heisenberg equations of motions. Section 3 analyzes the transmission of SC states, quantum interference effects with the help of reduced Wigner functions. The final section summarizes the results obtained in section 3.

\section{Heisenberg equations of motions}
We consider three-frequency interaction of light waves co-propagating (type-I) in a PPNC. The three interaction processes of light waves at frequencies  $\omega_{e}$, $2\omega_{e}$ and $3\omega_{e}$ are [10,11,14]:
\begin{eqnarray}
2\omega_{e}\rightarrow \omega_{e}+\omega_{e},\nonumber \\
\delta k_{1}=k_{2e}-2k_{1e}+m_{1}G_{1}=\Delta k_{1}+m_{1}G_{1},\\
2\omega_{e}+\omega_{e} \rightarrow 3\omega_{e},\nonumber \\
\delta k_{2}=k_{3e}-k_{2e}-k_{1e}+m_{2}G_{2}=\Delta k_{2}+m_{2}G_{2},
\end{eqnarray}
where $\Delta k_{s=1,2}$ is a phase mismatch for an ordinary nonlinear optical crystal for processes (1-2); $k_{se}$ is a wave number of interacting modes ($e$ stands for extraordinary wave); $G_{s}=2\pi\Lambda_{s}^{-1}$ is the modulus of the vector of a reciprocal lattice with period $\Lambda_{s}$; $m_{s}=\pm1,\pm3,...$ is the quasi-phase-matched order.\\

Equation (1) describes (parametric down conversion process), i.e. the splitting of a photon of frequency $2\omega_{e}$ into two photons of frequencies $\omega_{e}$. Further sum frequency generation process is described by (2), i.e., a photon of frequency $2\omega_{e}$ combines with the photon of frequency $\omega_{e}$, which creates a photon of frequency $3\omega_{e}$.\\

It has been shown that the nonlinear processes (1-2) can be quasi-phase-matched [10,11] in a single domain structure $(G_{1}=G_{2})$ or at coherence length $\approx 16 {\mu m}^{-1}$ when the light wave of frequency $\omega_{e}$ has wavelength $\approx 3.5 \mu m$ for PPLN crystal.
The processes (1-2) can be described by the nonlinear interaction Hamiltonian
\begin{eqnarray}
\hat{H}_{I}=\hbar f(z)[\xi_{1}\hat{a}_{1e}^{2}\hat{a}_{2e}^{+}e^{i\Delta{k_{1}z}}+\xi_{2}\hat{a}_{1e}\hat{a}_{2e}\hat{a}_{3e}^{+}e^{i\Delta{k_{2}z}}+HC].
\end{eqnarray}
Where $\hat{a}_{je} (\hat{a}_{je}^{+})$ are annihilation (creation) operators of photon of $je$th light wave at frequency $j\omega_{e}$; $\xi_{s}$ is the nonlinear coupling constant. Explicit form of $d_{eff}$ nonlinear coefficients and Sellmeier equations for considered nonlinear processes (1,2) in PPLN crystal can be found in [13,15]; $f(z)$ is the periodic function equal to $+1$ or $-1$ at domain thickness $l=\Lambda /2$. $HC$ denotes Hermitian conjugate. The operators $\hat{a}_{je}(\hat{a}_{je}^{+})$ obey the following commutation rules
\begin{eqnarray}[\hat{a}_{je},\hat{a}_{pe}^{+}]=\delta_{je,pe},  j,p=1,2,3.
\end{eqnarray}
The interaction Hamiltonian (3) can be averaged over the period $\Lambda$, if the interaction length $z$ is much more then period of modulation $\Lambda$, i.e. $z\gg \Lambda$. Then the interaction Hamiltonian (3) takes the form
\begin{eqnarray}
\hat{H}_{I}=\hbar[g_{1}\hat{a}_{1e}^{2}\hat{a}_{2e}^{+}+g_{2}\hat{a}_{1e}\hat{a}_{2e}\hat{a}_{3e}^{+}+HC],
\end{eqnarray}
where
\begin{eqnarray}
g_{s}=\int_{-\Lambda_{i}/2}^{+\Lambda_{i}/2}f(z)\exp(\pm i\Delta k_{s}z)dz=2\xi_{s}/(\pi m_{s}).
\end{eqnarray}
The Heisenberg operator equations corresponding to the interaction Hamiltonian (5) are given by
\begin{eqnarray}\label{e4}
\frac{d\hat{a}_{1e}}{dz}=\frac{1}{i\hbar}[\hat{a}_{1e},\hat{H}_{I}]=-ig_{3}\hat{a}^{+}_{2e}\hat{a}_{3e}-2ig_{2}\hat{a}^{+}_{1e}\hat{a}_{2e},\\
\label{e5}
\frac{d\hat{a}_{2e}}{dz}=\frac{1}{i\hbar}[\hat{a}_{2e},\hat{H}_{I}]=-ig_{2}\hat{a}^{2}_{1e}-ig_{3}\hat{a}^{+}_{1e}\hat{a}_{3e},\\
\label{e6}
\frac{d\hat{a}_{3e}}{dz}=\frac{1}{i\hbar}[\hat{a}_{3e},\hat{H}_{I}]=-ig_{3}\hat{a}_{1e}\hat{a}_{2e}.
\end{eqnarray}
Assuming the pump mode at frequency $2\omega_{e}$ is classical and non-depleted at the input of a PPNC
\begin{eqnarray}
\hat{a}_{2e}=A_{2}e^{i\phi_{2e}}.
\end{eqnarray}
After substitution of (10) into (7-9), we obtain following linear system of equations
\begin{eqnarray}\label{e8}
\frac{d\hat{a}_{1e}}{dz}=-ig_{2}A_{2}e^{-i\phi_{2e}}\hat{a}_{3e}-2ig_{1}A_{2}e^{i\phi_{2e}}\hat{a}^{+}_{1e},\\
\label{e9}
\frac{d\hat{a}_{3e}}{dz}=-ig_{2}A_{2}e^{i\phi_{2e}}\hat{a}_{1e}.
\end{eqnarray}
For simplicity, we introduce a normalized interaction length $\zeta$
\begin{eqnarray}\label{e10}
\zeta=zA_{2}g_{1}.
\end{eqnarray}
The quantity (13) is introduced into the set of equations (11-12), which reduce after straightforward algebra to the set of equations
\begin{eqnarray}
\frac{d\hat{a}_{1e}}{d\zeta}=-i\gamma e^{-i\phi_{2e}}\hat{a}_{3e}-2ie^{i\phi_{2e}}\hat{a}^{+}_{1e},\\
\frac{d\hat{a}_{3e}}{d\zeta}=-i\gamma e^{i\phi_{2e}}\hat{a}_{1e},
\end{eqnarray}
where $\gamma=g_{2}/g_{1}$. The set of linear equations (14-15) is solved by applying the Laplace transformation:
\begin{eqnarray}
\hat{a}_{1e}(\zeta)=\hat{a}_{1e}k_{1}(\zeta)+\hat{a}_{1e}^{+}k_{2}(\zeta)+\hat{a}_{3e}k_{3}(\zeta)+\hat{a}_{3e}^{+}k_{4}(\zeta),\\
\hat{a}_{3e}(\zeta)=\hat{a}_{1e}m_{1}(\zeta)+\hat{a}_{1e}^{+}m_{2}(\zeta)+\hat{a}_{3e}m_{3}(\zeta)+\hat{a}_{3e}^{+}m_{4}(\zeta),
\end{eqnarray}
where $\hat{a}_{je}=\hat{a}_{je}(0)$ and
\begin{eqnarray}
x_{1}=1+\sqrt{(1-\gamma^{2})},\nonumber \\
x_{2}=1-\sqrt{(1-\gamma^{2})},\nonumber \\
k_{1}(\zeta)=\frac{1}{(x_{2}^2-x_{1}^2)}[(x_{2}^{2}+\gamma^{2})\cosh{x_{2}\zeta}-(x_{1}^{2}+\gamma^{2})\cosh{x_{1}\zeta}],\nonumber \\
k_{2}(\zeta)=\frac{2ie^{i\phi_{2e}}}{(x_{2}^2-x_{1}^2)}[x_{1}\sinh{x_{1}\zeta}-x_{2}\sinh{x_{2}\zeta}],\nonumber \\
k_{3}(\zeta)=\frac{i\gamma e^{-i\phi_{2e}}}{2x_{1}x_{2}{(x_{2}^2-x_{1}^2)}}[x_{2}(\gamma^{2}+x_{1}^{2})\sinh{x_{1}\zeta}-x_{1}(\gamma^{2}+x_{1}^{2})\sinh{x_{2}\zeta}],\nonumber \\
k_{4}(\zeta)=\frac{2\gamma e^{2i\phi_{2e}}}{(x_{2}^2-x_{1}^2)}[\cosh{x_{2}\zeta}-\cosh{x_{1}\zeta}],\nonumber \\
m_{1}(\zeta)=\frac{
i\gamma e^{i\phi_{2e}}}{x_{1}x_{2}(x_{2}^2-x_{1}^2)}[x_{2}(\gamma^{2}+x_{1}^{2})\sinh{x_{1}\zeta}-x_{1}(\gamma^{2}+x_{2}^{2})\sinh{x_{2}\zeta}],\nonumber \\
m_{2}(\zeta)=\frac{2\gamma e^{2i\phi_{2e}}}{(x_{2}^2-x_{1}^2)}[\cosh{x_{1}\zeta}-\cosh{x_{2}\zeta}],\nonumber \\
m_{3}(\zeta)=\frac{1}{(x_{2}^2-x_{1}^2)}[(4-\gamma^{2}-x_{1}^{2})\cosh{x_{1}\zeta}-(4-\gamma^{2}-x_{2}^{2})\cosh{x_{2}\zeta}],\nonumber \\
m_{4}(\zeta)=\frac{i2\gamma^2 e^{3i\phi_{2e}}}{x_{1}x_{2}(x_{2}^2-x_{1}^2)}[x_{2}\cosh{x_{1}\zeta}-x_{1}\cosh{x_{2}\zeta}].
\end{eqnarray}
The solution (16-17) of the system of equations (14-15) can be verified by substituting expressions (18) into the commutation relations
\begin{eqnarray}
[\hat{a}_{1e}(\zeta),\hat{a}_{1e}^{+}(\zeta)]=k_{1}(\zeta)k_{1}^{*}(\zeta)-k_{2}(\zeta)k_{2}^{*}(\zeta)+k_{3}(\zeta)k_{3}^{*}(\zeta)-k_{4}(\zeta)k_{4}^{*}(\zeta)=1.
\end{eqnarray}
\begin{eqnarray}
[\hat{a}_{3e}(\zeta),\hat{a}_{3e}^{+}(\zeta)]=m_{1}(\zeta)m_{1}^{*}(\zeta)-m_{2}(\zeta)m_{2}^{*}(\zeta)+m_{3}(\zeta)m_{3}^{*}(\zeta)-m_{4}(\zeta)m_{4}^{*}(\zeta)=1.
\end{eqnarray}

\section{Composite Wigner quasi-probability distribution function}
One of the methods to visualize and analyze quantum interference effects [5-7,9] in quantum optics is to construct Wigner quasi-probability distribution function. With the help of Wigner function one can not only see the quantum interference effects but also evaluate quasi-probabilities for both Hermitian operators $\hat{x}$ and $\hat{p}$ on phase space simultaneously. In ordinary nonlinear optical crystals number of light waves interactions are limited but in PPNCs number of interactions [12,16] grow with the number of nonlinear processes (1-2). This growth of processes make calculations of Wigner function more complicated.\\

So, here we will first construct composite Wigner quasi-probability distribution function for light waves of frequencies $\omega_{e}$ and $3\omega_{e}$ and then after integration over either $\alpha_{1}$ or $\alpha_{3}$ will obtain reduced Wigner function of either $\alpha_{3}$ or $\alpha_{1}$. The composite quantum characteristic function $\hat{C}(\beta_{1},\beta_{3},\zeta)$ of light waves $\hat{a}_{1e}$ and $\hat{a}_{3e}$ can be written as
\begin{eqnarray}
\hat{C}(\beta_{1},\beta_{3},\zeta)=e^{\beta_{1}\hat{a}_{1e}^{+}(\zeta)-\beta_{1}^{*}\hat{a}_{1e}(\zeta)}e^{\beta_{3}\hat{a}_{3e}^{+}(\zeta)-\beta_{3}^{*}\hat{a}_{3e}(\zeta)}.
\end{eqnarray}

\subsection{Initial states}
The composite quantum characteristic function (21) is averaged over various initial states:
\subsubsection{Case I (Initial states: C in $\omega_{e}$, C in $3\omega_{e}$)}
$\hat{a}_{1e}$ and $\hat{a}_{3e}$ are in coherent states $|\alpha_{10},\alpha_{30}>$, i.e.
$|\varphi_{1}>=|\alpha_{10},\alpha_{30}>$;\\
\subsubsection{Case II (Initial states: C in $\omega_{e}$, V in $3\omega_{e}$)}
$\hat{a}_{1e}$ is in coherent state $|\alpha_{10}>$ and $\hat{a}_{3e}$ in vacuum state $|0_{30}>$, i.e. $|\varphi_{2}>=|\alpha_{10},0_{30}>$;\\
\subsubsection{Case III (Initial states: V in $\omega_{e}$, C in $3\omega_{e}$)}
 $\hat{a}_{1e}$ is in vacuum state $|0_{10}>$ and $\hat{a}_{3e}$ in coherent state $|\alpha_{30}>$, i.e. $|\varphi_{3}>=|0_{10},\alpha_{30}>$;\\
\subsubsection{Case IV (Initial states: V in $\omega_{e}$, V in $3\omega_{e}$)}
$\hat{a}_{1e}$ and $\hat{a}_{3e}$ are in vacuum states $|0_{10},0_{30}>$, i.e. $|\varphi_{4}>=|0_{10},0_{30}>$;\\
\subsubsection{Case V (Initial states: SC in $\omega_{e}$, V in $3\omega_{e}$)}
$\hat{a}_{1e}$ is in SC state $(|\alpha_{10}>+|-\alpha_{10}>)/\sqrt{2+2\exp(-2|\alpha_{10}|^{2})}$ and $\hat{a}_{3e}$ in vacuum state $|0_{30}>$, i.e. $|\varphi_{5}>=(|\alpha_{10}>+|-\alpha_{10}>)|0_{30}>/\sqrt{2+2\exp(-2|\alpha_{10}|^{2})}$;\\
\subsubsection{Case VI (Initial states: V in $\omega_{e}$, SC in $3\omega_{e}$)}
$\hat{a}_{1e}$ is in vacuum state $|0_{10}>$ and $\hat{a}_{3e}$ in SC state $(|\alpha_{30}>+|-\alpha_{30}>)/\sqrt{2+2\exp(-2|\alpha_{30}|^{2})}$, i.e. $|\varphi_{6}>=|0_{10}>(|\alpha_{30}>+|-\alpha_{30}>)/\sqrt{2+2\exp(-2|\alpha_{30}|^{2})}$;\\
\subsubsection{Case VII (Initial states: SC in $\omega_{e}$, C in $3\omega_{e}$)}
$\hat{a}_{1e}$ is in SC state $(|\alpha_{10}>+|-\alpha_{10}>)/\sqrt{2+2\exp(-2|\alpha_{10}|^{2})}$ and $\hat{a}_{3e}$ in coherent state $|\alpha_{30}>$, i.e. $|\varphi_{7}>=(|\alpha_{10}>+|-\alpha_{10}>)|\alpha_{30}>/\sqrt{2+2\exp(-2|\alpha_{10}|^{2})}$;\\
\subsubsection{Case VIII (Initial states: C in $\omega_{e}$, SC in $3\omega_{e}$)}
$\hat{a}_{1e}$ is in coherent state $|\alpha_{10}>$ and $\hat{a}_{3e}$ in SC state $(|\alpha_{30}>+|-\alpha_{30}>)/\sqrt{2+2\exp(-2|\alpha_{30}|^{2})}$, i.e. $|\varphi_{8}>=|\alpha_{10}>(|\alpha_{30}>+|-\alpha_{30}>)/\sqrt{2+2\exp(-2|\alpha_{30}|^{2})}$;\\
\subsubsection{Case IX (Initial states: SC in $\omega_{e}$, SC in $3\omega_{e}$)}
$\hat{a}_{1e}$ and $\hat{a}_{3e}$ are in SC states $(|\alpha_{10}>+|-\alpha_{10}>)(|\alpha_{30}>+|-\alpha_{30}>)/[\sqrt{2+2\exp(-2|\alpha_{10}|^{2})}\sqrt{2+2\exp(-2|\alpha_{30}|^{2})}]$, i.e. $|\varphi_{9}>=(|\alpha_{10}>+|-\alpha_{10}>)(|\alpha_{30}>+|-\alpha_{30}>)/[\sqrt{2+2\exp(-2|\alpha_{10}|^{2})}\sqrt{2+2\exp(-2|\alpha_{30}|^{2})}]$.\\

Composite Wigner function for light waves $\hat{a}_{1e}$ and $\hat{a}_{3e}$ can be obtained by taking Fourier transformation of quantum characteristic function (21)
\begin{eqnarray}
W_{p}(\alpha_{1},\alpha_{3})=\frac{1}{\pi^{3}}\int \int \int C_{p}(\beta_{1},\beta_{3},\zeta)e^{\beta_{1}^{*}\alpha_{1}-\beta_{1}\alpha_{1}^{*}}e^{\beta_{3}^{*}\alpha_{3}-\beta_{3}\alpha_{3}^{*}}d^{2}\beta_{1}d^{2}\beta_{3},
\end{eqnarray}
where $C_{p}(\beta_{1},\beta_{3},\zeta)=Tr[\hat{\rho}_{p}\hat{C}(\beta_{1},\beta_{3},\zeta)]$, $\hat{\rho}_{p}=|\varphi_{p}><\varphi_{p}|$, and $p=1,2,..9$.\\

The reduced Wigner function for light wave $\hat{a}_{1e}$ is obtained by
\begin{eqnarray}
W_{p}(\alpha_{1})=\frac{1}{\pi}\int W_{p}(\alpha_{1},\alpha_{3})d^{2}\alpha_{3},
\end{eqnarray}

and for light wave $\hat{a}_{3e}$
\begin{eqnarray}
W_{p}(\alpha_{3})=\frac{1}{\pi}\int W_{p}(\alpha_{1},\alpha_{3})d^{2}\alpha_{1}
\end{eqnarray}

The reduced Wigner function (23-24) is analytically calculated for cases $I-IX$ and their visualization demonstrates well known nonclassical properties of light waves of frequencies $\omega_{e}$ and $3\omega_{e}$ in a PPNC [12].

\subsection{Phase-space 'portraits'}
'Portraits' of transmission of superposition of macroscopically distinguishable states and interference between them are clearly visible on phase space representation.

\subsubsection{Case I (Initial states: C in $\omega_{e}$, C in $3\omega_{e}$)}
First squeezing of light wave of frequency $\omega_{e}$ begins and then it goes to $3\omega_{e}$ and later vice versa.
\subsubsection{Case II (Initial states: C in $\omega_{e}$, V in $3\omega_{e}$)}
Behavior of squeezing effects for this case is very similar to the effect in case I.
\subsubsection{Case III (Initial states: V in $\omega_{e}$, C in $3\omega_{e}$)}
Behavior of squeezing effects for this case is very similar to the effect in case I.
\subsubsection{Case IV (Initial states: V in $\omega_{e}$, V in $3\omega_{e}$)}
Behavior of squeezing effects for this case is very similar to the effect in case I.
\subsubsection{Case V (Initial states: SC in $\omega_{e}$, V in $3\omega_{e}$)}
Figures 1a and 1b demonstrate transmission of SC state from light wave of frequency $\omega_{e}$ into $3\omega_{e}$. Quantum interference between superposition of macroscopically distinguishable states are well seen. This is due to the nonlinear processes (1-2) and correlation effect [12,14] present in between light waves of frequencies $\omega_{e}$ and $3\omega_{e}$.
\subsubsection{Case VI (Initial states: V in $\omega_{e}$, SC in $3\omega_{e}$)}
Behavior of transmission, interference and squeezing effects are similar to the effects in case $V$. But here transmission process as compared to case $V$ at earlier stages of interaction flows from $3\omega_{e}$ to $\omega_{e}$ and later starts competing with $\omega_{e}$ to $3\omega_{e}$.
\subsubsection{Case VII (Initial states: SC in $\omega_{e}$, C in $3\omega_{e}$)}
Behavior of transmission, interference and squeezing effects are similar to the case $V$.
\subsubsection{Case VIII (Initial states: C in $\omega_{e}$, SC in $3\omega_{e}$)}
Behavior of transmission, interference and squeezing effects are similar to the case $VI$.
\subsubsection{Case IX (Initial states: SC in $\omega_{e}$, SC in $3\omega_{e}$)}
At earlier stages of interaction figures 2a and 2b demonstrate transmission and formation of superposition of 4 macroscopically distinguishable states in light waves of frequencies $\omega_{e}$ and $3\omega_{e}$. Phase space 'portraits' (fig. 2a and 2b) illustrate quantum interferences between superposition of 4 states. Moreover, all 4 states are in squeezed states. At later stages of interaction superposition of 4 macroscopically distinguishable states becomes more complicated in light wave of frequency $3\omega_{e}$.
\subsubsection{Case X: Interference effect. Initial states: V in $\omega_{e}$, SC in $3\omega_{e}$}
Figures 53 and 3b demonstrate how to manage interference effect present in between superposition of two macroscopically distinguishable states by selecting values of nonlinear effective coefficient $d_{eff}$ [13].
\section{Conclusion}
In this paper we have investigated the transmission of SC states and interference effects in a PPNC with second order nonlinear susceptibility. In the case of classical and non-depleted wave of frequency $2\omega_{e}$ at the input of a PPNC SC, C and V states are studied with the help of reduced Wigner function. Phase-space 'portraits' of reduced Wigner function for light wave of frequencies $\omega_{e}$ and $3\omega_{e}$ are used to visualize the transmission of SC states. It is shown that the quantum interference, quantum transmission of SC states can be achieved in a PPNC.

In cases $V-VI$ and $IX$, we have shown that the transmission of SC can be achieved in both directions, i.e. from $\omega_{e}$ to $3\omega_{e}$ and vice versa. In case $VI$, it is shown that the quantum interference between 2 superposition of macroscopically distinguishable states can be managed (case $X$) by selecting appropriate values of nonlinear $d_{eff}$ coefficients [13].\\

So, PPNCs can be good candidates for quantum state engineering. With the help of PPNCs one can transmit superposition of macroscopically distinguishable states and such delicate effects like quantum interfaces and squeezing from one light wave to other.

\section{Acknowledgements}
I am grateful to Prof. A. S. Chirkin, Prof. S. P. Kulik, Prof. V. N. Zadkov, Dr. D. N. Yanyshev, Prof. V. P. Kandidov, and Dr. E. P. Silaeva for providing invaluable support.

\newpage

\begin{figure}
\includegraphics[height=10cm,width=10cm]{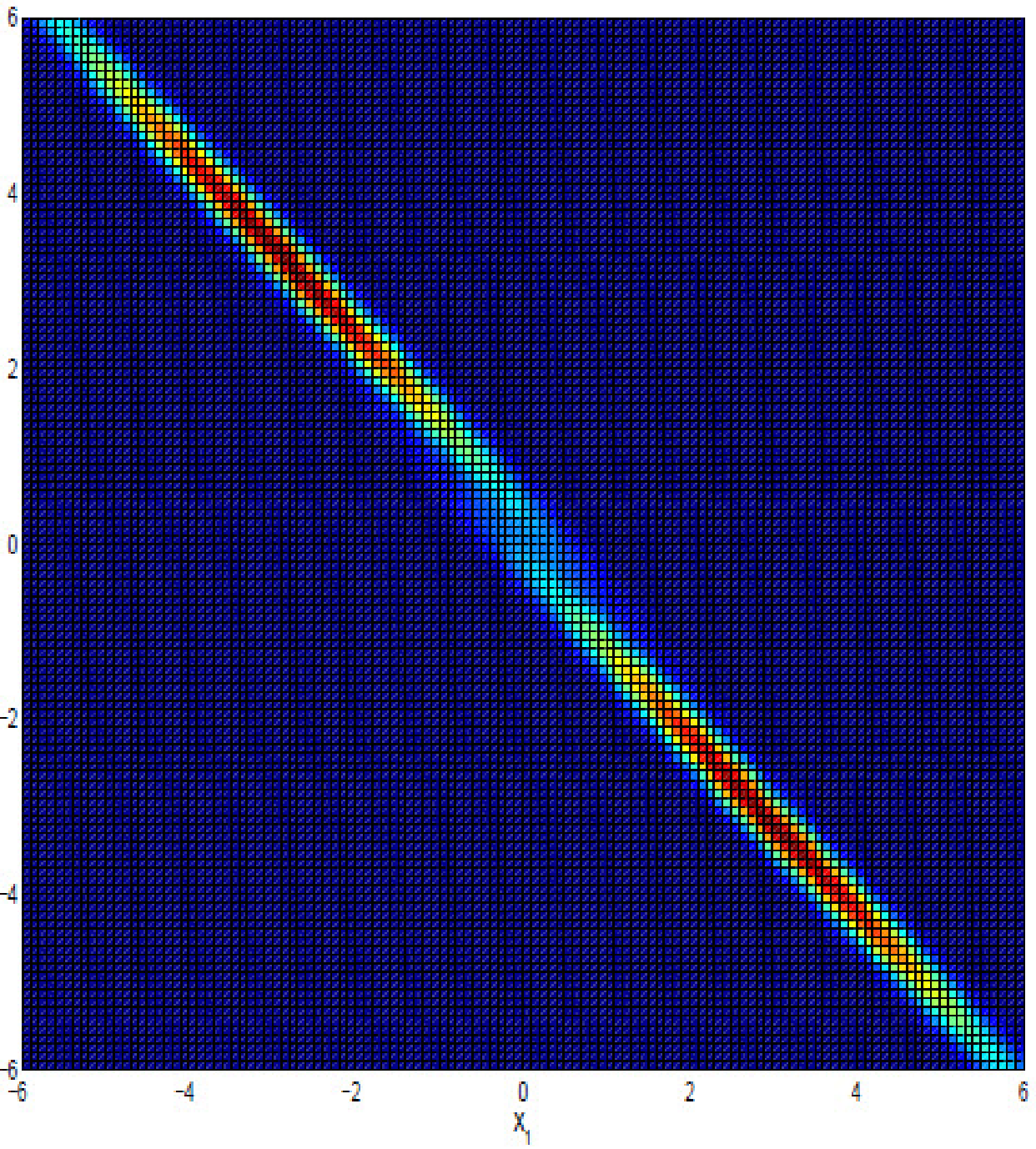}
\caption{
Fig. 1a. Phase space 'portrait' of $W_{5}(\alpha_{1})$, when $\phi_{2e}=0$, average number of photons in the wave of frequency $\omega_{e}$, $|\alpha_{10}|^{2}=12$, $\alpha_{10}=\sqrt{12}e^{i\pi /3}$, and the wave of frequency $3\omega_{e}$, $|\alpha_{30}|^{2}=0$, $\alpha_{30}=\sqrt{0}e^{i\phi_{3e}}$, normalized coherence length $\zeta=0.9$, and normalized nonlinear coupling constant $\gamma=0.9$.}
\end{figure}

\begin{figure}
\includegraphics[height=10cm,width=10cm]{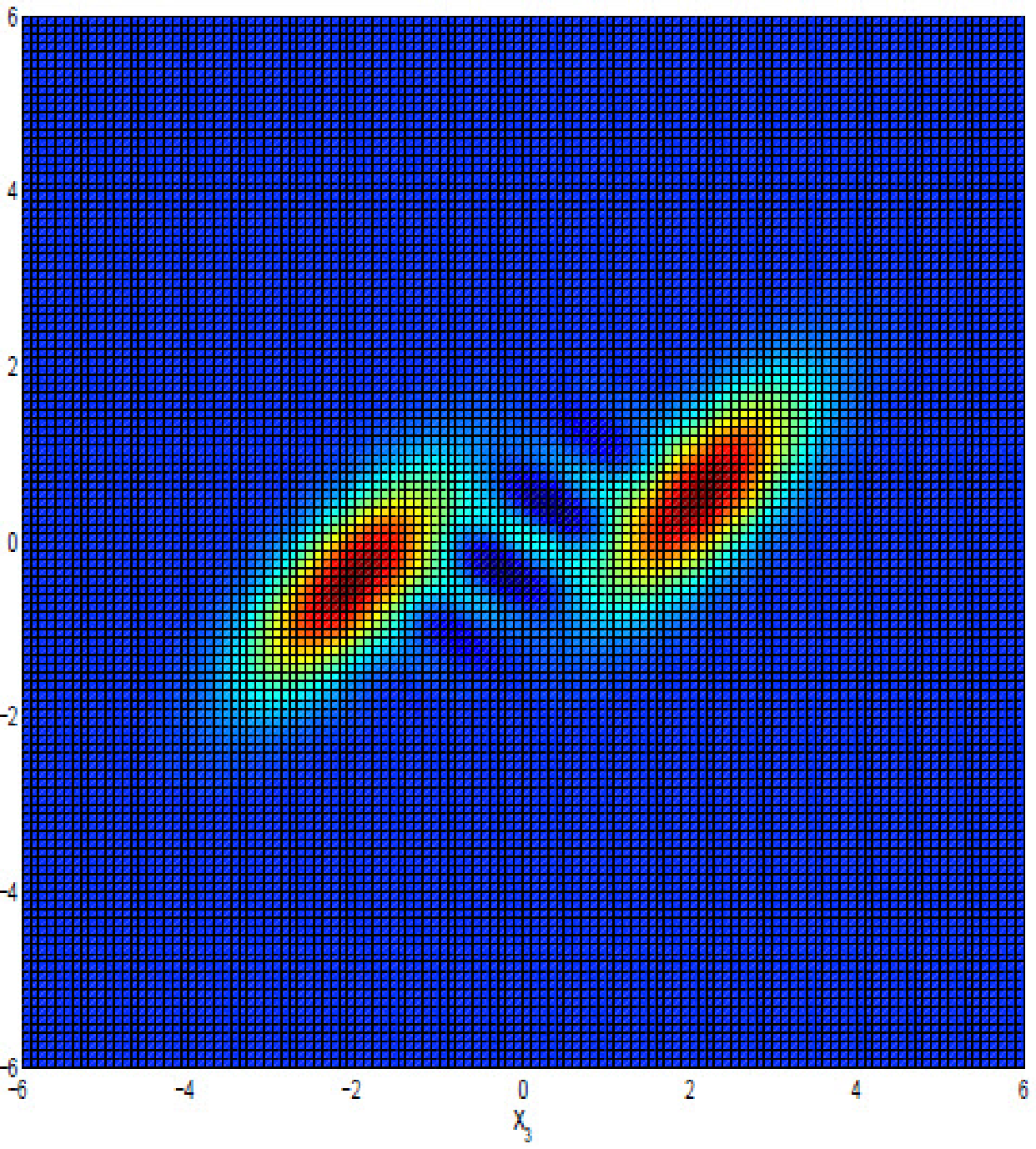}
\caption{
Fig. 1b. Phase space 'portrait' of $W_{5}(\alpha_{3})$, when $\phi_{2e}=0$, average number of photons in the wave of frequency $\omega_{e}$, $|\alpha_{10}|^{2}=12$, $\alpha_{10}=\sqrt{12}e^{i\pi /3}$, and the wave of frequency $3\omega_{e}$, $|\alpha_{30}|^{2}=0$, $\alpha_{30}=\sqrt{0}e^{i\phi_{3e}}$, normalized coherence length $\zeta=0.9$, and normalized nonlinear coupling constant $\gamma=0.9$.}
\end{figure}

\begin{figure}
\includegraphics[height=10cm,width=10cm]{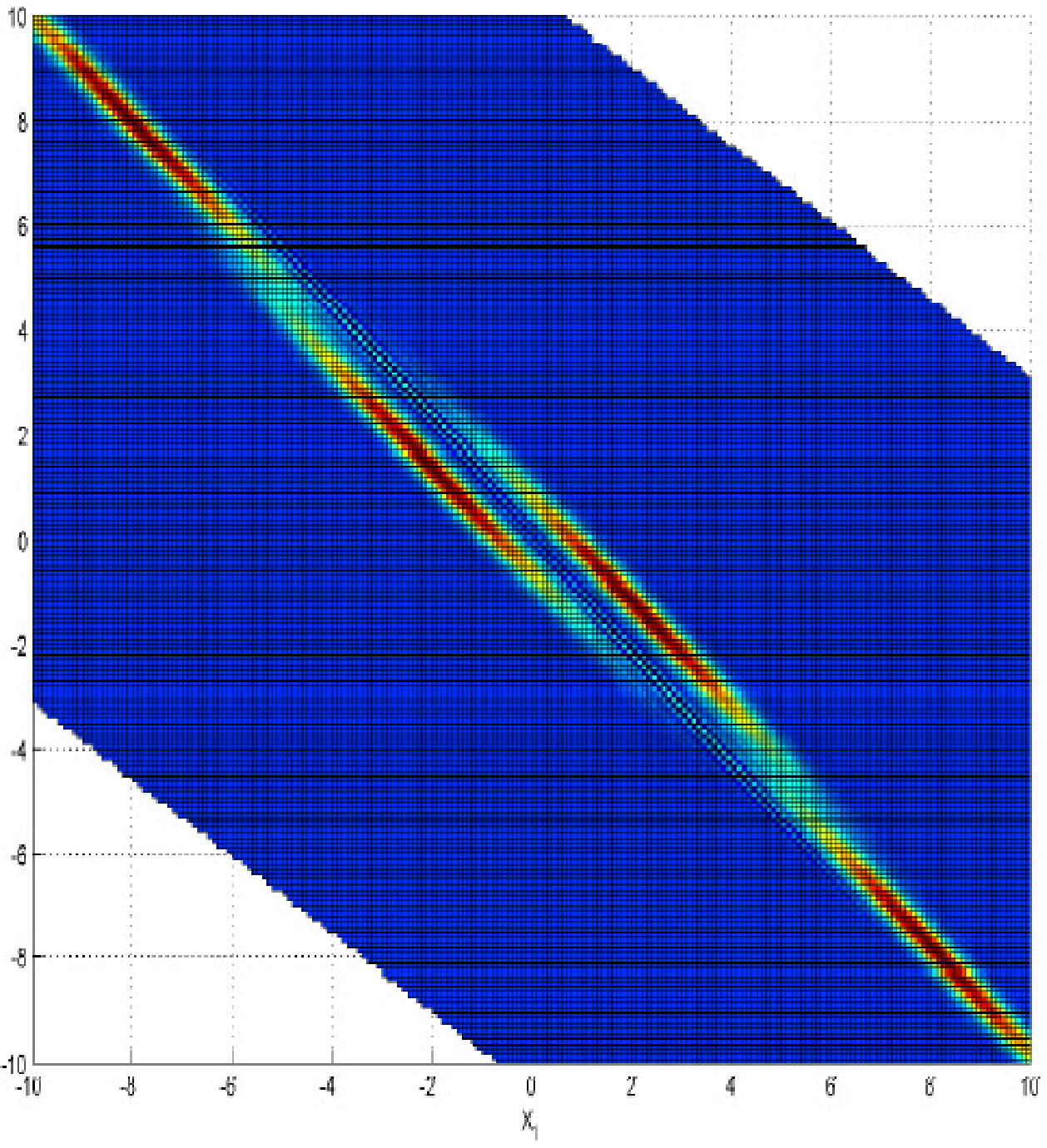}
\caption{
Fig. 2a. Phase space 'portrait' of $W_{9}(\alpha_{1})$, when $\phi_{2e}=0$, average number of photons in the wave of frequency $\omega_{e}$, $|\alpha_{10}|^{2}=12$, $\alpha_{10}=\sqrt{12}e^{i\pi /3}$, and the wave of frequency $3\omega_{e}$, $|\alpha_{30}|^{2}=12$, $\alpha_{30}=\sqrt{12}e^{i\pi /3}$, normalized coherence length $\zeta=0.9$, and normalized nonlinear coupling constant $\gamma=0.9$.}
\end{figure}

\begin{figure}
\includegraphics[height=10cm,width=10cm]{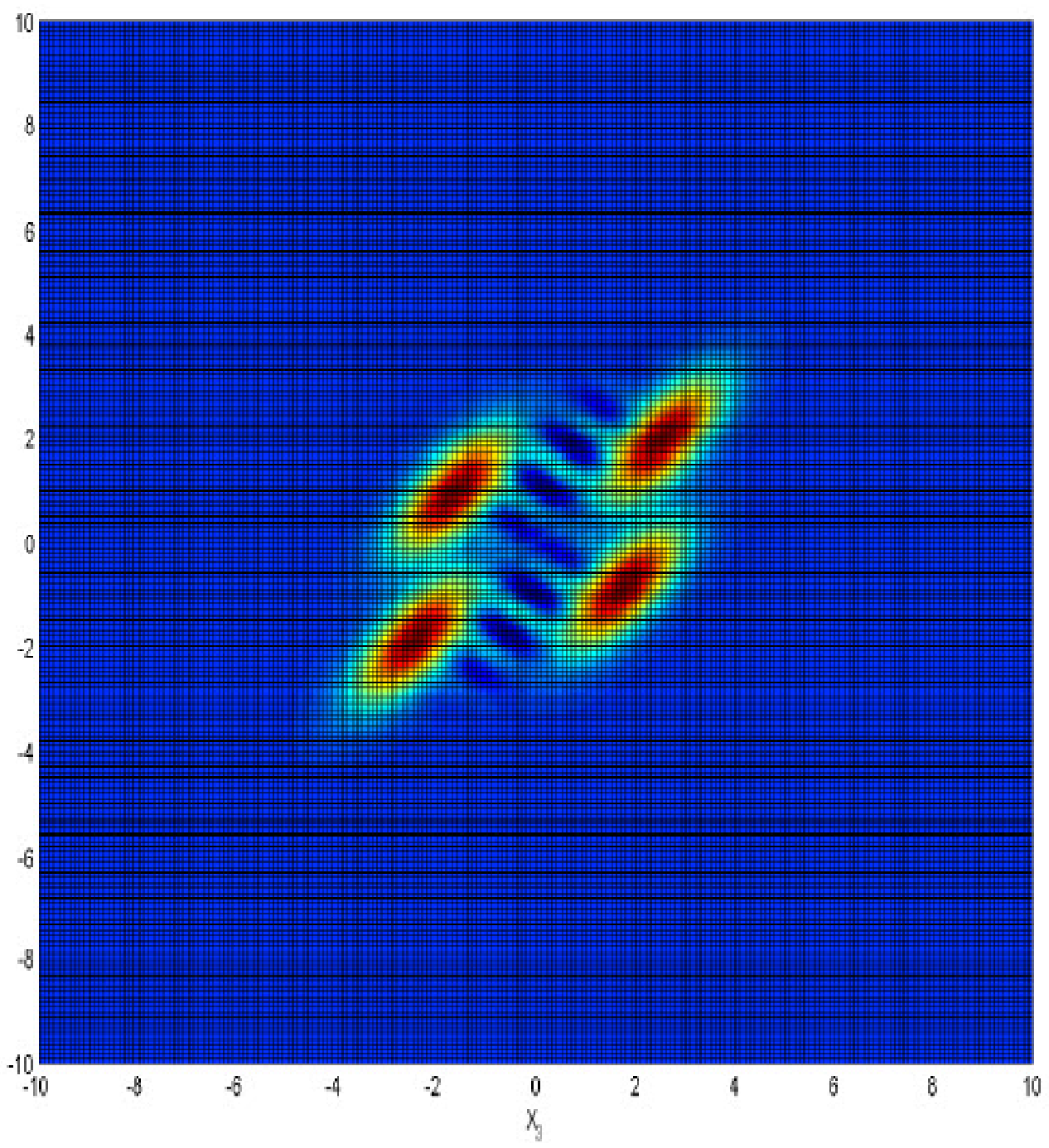}
\caption{
Fig. 2b. Phase space 'portrait' of $W_{9}(\alpha_{3})$, when $\phi_{2e}=0$, average number of photons in the wave of frequency $\omega_{e}$, $|\alpha_{10}|^{2}=12$, $\alpha_{10}=\sqrt{12}e^{i\pi /3}$, and the wave of frequency $3\omega_{e}$, $|\alpha_{30}|^{2}=12$, $\alpha_{30}=\sqrt{12}e^{i\pi /3}$, normalized coherence length $\zeta=0.9$, and normalized nonlinear coupling constant $\gamma=0.9$.}
\end{figure}

\begin{figure}
\includegraphics[height=10cm,width=10cm]{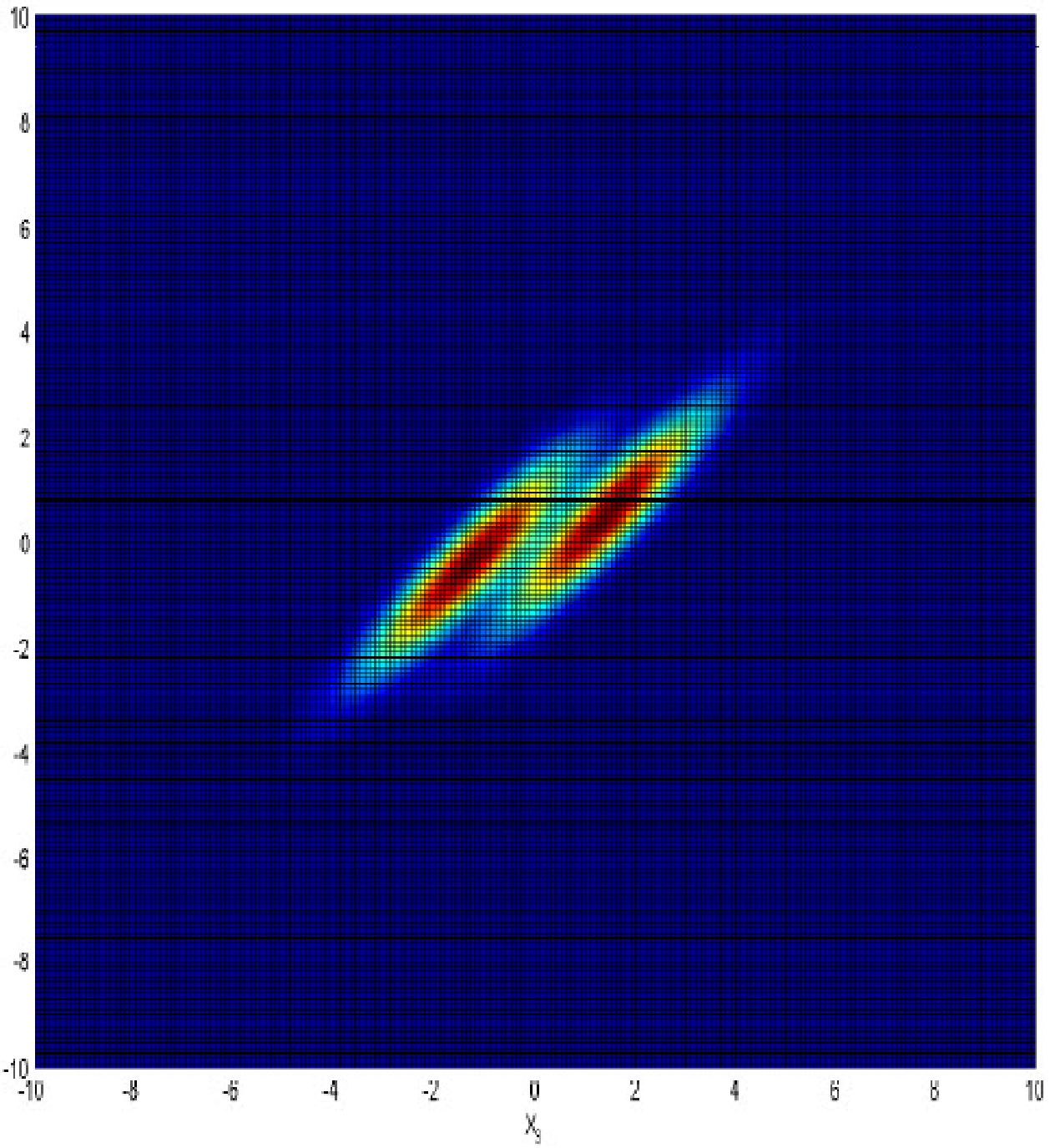}
\caption{
Fig. 3a. Phase space 'portrait' of $W_{6}(\alpha_{3})$, when $\phi_{2e}=0$, average number of photons in the wave of frequency $\omega_{e}$, $|\alpha_{10}|^{2}=0$, $\alpha_{10}=\sqrt{0}e^{0i\phi_{1e}}$, and the wave of frequency $3\omega_{e}$, $|\alpha_{30}|^{2}=12$, $\alpha_{30}=\sqrt{12}e^{i\pi /3}$, normalized coherence length $\zeta=1.2$, and normalized nonlinear coupling constant $\gamma=0.9$}
\end{figure}

\begin{figure}
\includegraphics[height=10cm,width=10cm]{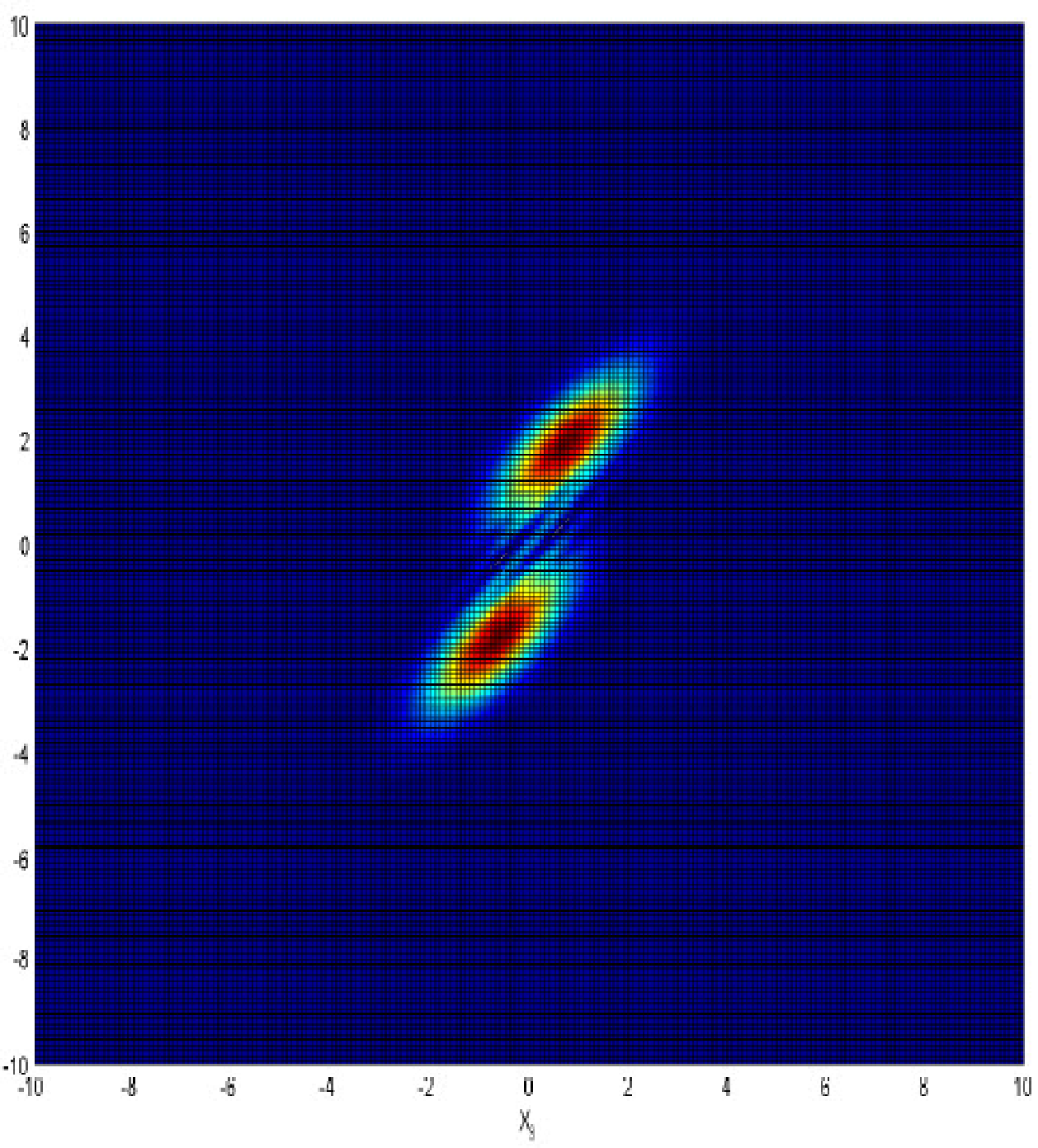}
\caption{
Fig. 3b. Phase space 'portrait' of $W_{6}(\alpha_{3})$, when $\phi_{2e}=0$, average number of photons in the wave of frequency $\omega_{e}$, $|\alpha_{10}|^{2}=0$, $\alpha_{10}=\sqrt{0}e^{0i\phi_{1e}}$, and the wave of frequency $3\omega_{e}$, $|\alpha_{30}|^{2}=12$, $\alpha_{30}=\sqrt{12}e^{i\pi /3}$, normalized coherence length $\zeta=1.2$, and normalized nonlinear coupling constant $\gamma=0.5$.}
\end{figure}
\end{document}